\documentclass[aps,prd,superscriptaddress,notitlepage,12pt]{revtex4-1}

\begin{document}
\title{Examination of a simple example of gravitational wave memory}

\author{Alexander Tolish}
\email{tolish@uchicago.edu}
\affiliation{Enrico Fermi Institute and Dept. of Physics, University of Chicago, Chicago, IL, 60637, USA}
\author{Lydia Bieri}
\email{lbieri@umich.edu}
\affiliation{Dept. of Mathematics, University of Michigan, Ann Arbor, MI 48109-1120, USA}
\author{David Garfinkle}
\email{garfinkl@oakland.edu}
\affiliation{Dept. of Physics, Oakland University, Rochester, MI 48309, USA}
\affiliation{Michigan Center for Theoretical Physics, Randall Laboratory of Physics, University of Michigan, Ann Arbor, MI 48109-1120, USA}
\author{Robert M. Wald}
\email{rmwa@uchicago.edu}
\affiliation{Enrico Fermi Institute and Dept. of Physics, University of Chicago, Chicago, IL, 60637, USA}

\date{\today}

\begin{abstract}
We examine a simple example of gravitational wave memory due to the decay of a point particle into two point particles.  In the case where one of the decay products is null, there are two types of memory: a null memory due to the null particle and an ordinary memory due to the recoiling timelike particle.  In the case where both decay products are timelike, there is only ordinary memory.  However, this ordinary memory can mimic the null memory in the limit where one of the decay products has a large velocity.  

\end{abstract}


\maketitle

\section{Introduction}
\indent

Gravitational wave memory, a permanent change in the detector after the gravitational wave has passed, has been known since
the work of Zel'dovich and Polnarev.\cite{zeldovich}  In the weak field slow motion approximation there is a simple relation between the gravitational wave memory and the net change in the second time derivative of the quadrupole moment of the source.  However, the simple picture of memory based on the weak field slow motion case was changed by the work of Christodoulou\cite{christodoulou} who found an additional memory effect related to the energy carried away in gravitational radiation.  The work of \cite{christodoulou} (based on the stability result of Christodoulou and Klainerman\cite{globalCK}) used the full non-linear Einstein equation, and presented the additional memory effect as due to the non-linear treatment, calling the previously known memory effect ``linear'' and the new memory ``nonlinear.''  However, in cases where electromagnetic 
waves\cite{zipser,lydia1,lydia2} or neutrinos\cite{epstein,turner,nullfluid} are present, they also contribute to the ``nonlinear'' memory effect with the energy of their radiation playing exactly the same role as that of the energy of gravitational waves.  Because of this, it has recently been argued \cite{lydiaandme} that instead of ``linear'' and ``nonlinear'' one should think of these two types of memory as ``ordinary'' and ``null'' where the null memory is due to stress-energy that gets out to null infinity.  

The weak field slow motion approximation allows the construction of many simple examples of sources whose gravitational wave memory can be calculated explicitly.  However, it is more difficult to find a simple, explicit example of the null memory.  Recently such an example was provided in \cite{twmemory} which calculates the memory due to the gravitational waves created by the decay of a particle of mass $M$ that emits a null particle of energy $E$.  The decay process gives rise to an impulsive gravitational wave whose Weyl tensor can be calculated explicitly.  The memory is then calculated by integrating the geodesic deviation equation to find the permanent displacement of the gravitational wave detector.  In \cite{twmemory} it is assumed that $E \ll M$ and the memory is calculated to first order in $E/M$.  The results are shown to agree with the formulas for the null memory given in \cite{christodoulou,lydiaandme}.  This agreement is both reassuring and puzzling.  The puzzling part comes from the fact that there are two types of memory treated in \cite{lydiaandme}: ordinary memory and null memory.  So why does the calculation of \cite{twmemory} agree with just the null part of the memory?  One possible explanation is that the calculation of \cite{twmemory} is first order in $E/M$ and the ordinary memory occurs only at higher order.  To see whether this explanation is correct, we will in section II extend the calculation of \cite{twmemory} by finding the memory due to the decay process without making any assumption of smallness of $E/M$.  We will then compare to the memory formula of \cite{lydiaandme} to see how much of the memory is ``ordinary'' and how much of the memory is ``null.''

In section III we will generalize the calculation of \cite{twmemory} in a different way.  Once again the particle of mass $M$ will decay by emitting a particle of energy $E$; however in this case the emitted particle will travel at a speed $\beta < 1$ rather than at the speed of light.  We will calculate the memory both by finding the impulsive gravitational wave using the method of \cite{twmemory} and by using the memory formula of \cite{lydiaandme}.  Here the memory will only be ordinary memory, since the emitted particle is traveling at a speed slower than light.  Nonetheless, in the limit as $\beta \to 1$ we would expect to recover the results of the previous section.  That is, we would expect the ordinary memory of the emitted particle to somehow ``imitate'' the null memory of the null particle with that imitation becoming better as the speed of the emitted particle approaches the speed of light.  By performing the calculation we will see to what extent and in what sense this expectation is correct.  

Our conclusions are given in section IV.       

\section{memory of a null particle}

In calculations of gravitational wave memory, the ends of the detector are assumed to travel on geodesics, and the memory is
essentially the second integral of the geodesic deviation equation.  In perturbation theory, the metric is written as 
${g_{ab}}={\eta_{ab}}+{h_{ab}}$ where ${\eta_{ab}}$ is a flat metric and $h_{ab}$ is small.  The geodesic deviation equation
governing the separation $D^a$ of two nearby geodesics is
\begin{equation}
{{{d^2}{D_a}} \over {d {t^2}}} = - {R_{tatb}} {D^b} \; \; \; ,
\label{geodev}
\end{equation}
where the Riemann tensor $R_{abcd}$ is given in terms of the metric perturbation by 
\begin{equation}
{R_{abcd}} = {\textstyle {\frac 1 2}} \left ( {\partial _b}{\partial _c}{h_{ad}} + {\partial _a}{\partial _d}{h_{bc}} - {\partial _b}{\partial _d}{h_{ac}} - {\partial _a}{\partial _c}{h_{bd}}\right ) \; \; \; .
\label{Riemann}
\end{equation}  
Since the detector is assumed to be far from the source, one only needs the Riemann tensor, and therefore the metric perturbation, to order $1/r$.

We now specialize to the particular case treated in \cite{twmemory}: the decay, at time $t=0$, of a particle of mass $M$ at rest into a null particle of energy $E$ that travels in the $z$ direction and a recoiling particle of mass $M'$ that travels in the $-z$ direction.  Due to the axisymmetry of the problem, it follows that to order $1/r$ we have
\begin{equation}
{R_{tatb}} = W ({\theta_a} {\theta_b} - {\phi_a}{\phi_b}) \; \; \; ,
\label{electricWeyl}
\end{equation}
for some scalar $W$.  Here $\theta ^a$ and $\phi ^a$ are unit vectors in the $\theta$ and $\phi$ directions. From eqn. (\ref{electricWeyl}) it follows that we only need to calculate one component of $R_{tatb}$ to find all components.  In particular, it follows from the standard expressions for spherical coordinates that
\begin{equation}
W = {{R_{txty}}\over {(1+{\cos^2}\theta)\cos\phi \sin\phi}} \; \; \; .
\label{Wcalc}
\end{equation}
The reason for choosing $R_{txty}$ is that for the metric of \cite{twmemory}, this component is particularly
simple to calculate.  Since ${h_{xy}}, \, {h_{tx}}$ and $h_{ty}$ all vanish, it follows from eqn.(\ref{Riemann}) that 
\begin{equation}
{R_{txty}} = - {\textstyle {1\over 2}} {\partial _x}{\partial _y} {h_{tt}} \; \; \; ,
\label{WeylExy1}
\end{equation}
so the only component of the metric that we need is $h_{tt}$.  However to leading order in $1/r$ it follows from eqns. (44-46)
of \cite{twmemory} that 
\begin{equation}
{h_{tt}} = {k\over r} + {1\over r} \Theta (u) \left [ - 2 M + {{2 {M '}\gamma (1+{v^2})} \over {(1 + v \cos \theta)}} + {{4E}\over {1 - \cos \theta}} \right ] \; \; \; .
\label{htt1}
\end{equation}
Here $k$ is a constant, $\Theta$ is the step function, $u=t-r$ is the retarded time, and $v$ and $\gamma$ are respectively the speed and gamma factor of the recoiling particle.  Note however that $M '$ and $v$ are not independent quantities.  Rather they are determined by $M$ and $E$ as follows: The conservation of four-momentum for the decay is
\begin{eqnarray}
{M '} \gamma = M - E \; \; \; ,
\label{conserveE}
\\
{M '} \gamma v = E \; \; \; .
\label{conserveP}
\end{eqnarray}
Using eqns. (\ref{conserveE}-\ref{conserveP}) in eqn. (\ref{htt1}), some straightforward but tedious algebra yields
\begin{equation}
{h_{tt}} = {k\over r} + {2\over r} \Theta (u)  {{E (1 + {\cos ^2}\theta)} \over {(1 - \cos \theta )(1 - (E/M)(1 - \cos \theta))}} \; \; \; .
\label{htt2}
\end{equation}
Then using eqn. (\ref{htt2}) in eqn. (\ref{WeylExy1}) we obtain
\begin{equation}
{R_{txty}} = {{-1}\over r} {\delta '} (u)  {{E (1 + {\cos ^2}\theta)} \over {(1 - \cos \theta )(1 - (E/M)(1 - \cos \theta))}} {\sin ^2} \theta \cos \phi \sin \phi \; \; \; .
\label{WeylExy2}
\end{equation}
Here $\delta$ is the Dirac delta function.
Now using eqn. (\ref{WeylExy2}) in eqn. (\ref{Wcalc}) we obtain
\begin{equation}
W = {{- 1}\over r} {\delta '} (u) {{E(1 + \cos \theta)} \over {1 - (E/M)(1 - \cos \theta)}} \; \; \; .
\label{Wresult}
\end{equation}
Finally, using eqn. (\ref{Wresult}) in eqn. (\ref{geodev}) and integrating twice with respect to time, we find that the
net change in separation $\Delta {D_a}$ is given by
\begin{equation}
\Delta {D_a} = {\frac E r} {\frac {1+\cos \theta} {1-(E/M)(1-\cos \theta )}} ({\theta _a}{\theta _b} - {\phi _a}{\phi _b}) {D^b} \; \; \; .
\label{memoryresult}
\end{equation}

We now compare the result of eqn. (\ref{memoryresult}) to the memory formula of \cite{lydiaandme}.  In particular, we would like to know how much of the memory is ordinary memory and how much is null memory.  In \cite{lydiaandme} the memory is expressed in terms of a tensor $m_{AB}$ on the unit two-sphere.  (Here capital letter indicies are used for tensors on the unit two-sphere, and all such indicies are raised and lowered with the unit two-sphere metric).  Given an initial separation $d$ in the $B$ direction, the change in separation in the $A$ direction is given by
\begin{equation}
\Delta d = - {\frac d r} {{m^A}_B} \; \; \; .
\end{equation}
To make comparisons with the notation and results of \cite{twmemory} note that in the case of axisymmetry the memory tensor
takes the form
\begin{equation}
{m_{AB}} = C(\theta) ({\theta_A}{\theta_B}-{\phi_A}{\phi_B}) \; \; \; ,
\label{aximemory}
\end{equation}  
for some function $C(\theta )$.  Here $\theta_A$ and $\phi _A$ are unit vectors on the unit two-sphere in the 
$\theta$ and $\phi$ directions.  It then follows that the net change in separation can be expressed as
\begin{equation}
\Delta {D_a} = - {\frac {C(\theta )} r} ({\theta _a}{\theta _b} - {\phi _a}{\phi _b}) {D^b} \; \; \; .
\label{memoryformula}
\end{equation}  
The memory tensor is determined as the solution of the following system:
\begin{eqnarray}
{D_A}{D^A}\Phi = \Delta P - 8 \pi F \; \; \; ,
\label{laplace}
\\
{D^B}{m_{AB}} = {D_A}\Phi \; \; \; .
\label{divm}
\end{eqnarray}
Here $D_A$ is the derivative operator on the unit two sphere and $F$ is the energy per unit solid angle radiated to null infinity.  The quantity $\Delta P$ is defined as follows: define $P(u,\theta ,\phi )$ to be the limit to null infinity of
${r^3}{C_{trtr}}$ where $C_{abcd}$ is the Weyl tensor.  Then  $P(\pm \infty)$ is defined to be 
${\lim _{u \to \pm \infty}} P$ and $\Delta P (\theta,\phi) $ is defined to be $P(\infty)-P(-\infty)$.  Due to the axisymmetry of the problem, there must be functions $A(\theta)$ and $B(\theta)$ such that
\begin{eqnarray}
\Delta P - 8 \pi F = A(\theta) \; \; \; ,
\label{Adef}
\\
\Phi = B(\theta) \; \; \; .
\label{Bdef}
\end{eqnarray} 
The consistency of eqns. (\ref{laplace}-\ref{divm}) requires that $A(\theta)$ have vanishing $\ell =0$ and $\ell =1$ part.  

Using the ansatz of eqns. (\ref{aximemory},\ref{Adef},\ref{Bdef} ) we find that eqns. (\ref{laplace},\ref{divm}) become
\begin{eqnarray}
{d \over {d\theta}} \left ( \sin \theta {{dB}\over {d\theta}}\right ) = \sin \theta A \; \; \; ,
\label{axilaplace}
\\
{d \over {d\theta}} \left ( {\sin^2} \theta C\right ) = {\sin^2} \theta {{dB}\over {d\theta}} \; \; \; .
\label{axidivm}
\end{eqnarray}
The memory can be divided into ordinary memory and null memory as follows:  $\Phi={\Phi_1}+{\Phi_2}$ which satisfy
\begin{eqnarray}
{D_A}{D^A}{\Phi_1} = \Delta P - {{(\Delta P)}_{[1]}} \; \; \; ,
\\
{D_A}{D^A}{\Phi_2} = - 8 \pi (F - {F_{[1]}}) \; \; \; ,
\label{laplacenull}
\end{eqnarray}
where the subscript $[1]$ denotes the $\ell =0$ and $\ell =1$ part.  Then ${m_{AB}}={m_{1AB}}+{m_{2AB}}$ which satisfy
\begin{eqnarray}
{D^B}{m_{1AB}} = {D_A}{\Phi_1} \; \; \; ,
\\
{D^B}{m_{2AB}} = {D_A}{\Phi_2} \; \; \; .
\label{divmnull}
\end{eqnarray}
Here $m_{1AB}$, the memory due to $\Delta P$, is the ordinary memory, while $m_{2AB}$, the memory due to $F$, is the null memory.  In each case, ordinary memory or null memory, eqns. (\ref{axilaplace}-\ref{axidivm}) hold, with 
$A(\theta) = \Delta P - {{(\Delta P)}_{[1]}}$in the case of ordinary memory and 
$A(\theta) =  - 8 \pi (F - {F_{[1]}})$ in the case of null memory.

We now work out the null memory for the case of the decay of a particle of mass $M$ emitting a null particle of energy $E$.  Since the particle is emitted in the $z$ direction, it follows that $F = E \delta$ where $\delta$ is the delta function which vanishes everywhere except $\theta =0$ and whose integral over the unit two sphere is 1.  We then find
\begin{equation}
- 8 \pi (F - {F_{[1]}}) = 2E (- 4 \pi \delta + (1+ 3 \cos \theta)) \; \; \; .
\label{nullsource}
\end{equation}
Thus to find the null memory, we must solve eqns. (\ref{axilaplace}-\ref{axidivm}) with $A$ given by the right hand side
of eqn. (\ref{nullsource}).  Note that since $\delta$ vanishes for $\theta > 0$, what we need to do is to solve 
eqns. (\ref{axilaplace}-\ref{axidivm}) for $\theta > 0$ with $A = 2 E (1+3\cos \theta)$.  Given a solution for 
$\theta > 0$ we can then verify that eqns. (\ref{laplace}-\ref{divm}) are satisfied in a distributional sense.  We will solve the equations in this section and then give the demonstration that the solution is a distributional solution in 
Appendix \ref{distribution}. 
For $\theta > 0$ eqn. (\ref{axilaplace}) becomes
\begin{equation}
{d \over {d\theta}} \left ( \sin \theta {{dB}\over {d\theta}}\right ) = E (2 \sin \theta + 6 \sin \theta \cos \theta ) \; \; \; ,
\end{equation}
from which we find
\begin{equation}
\sin \theta {{dB}\over {d\theta}} = E (- 2 \cos \theta + 3 {\sin ^2} \theta + {c_0}) \; \; \; ,
\end{equation}
for some constant $c_0$.  Since the left hand side of this equation vanishes at $\theta = \pi$, we must have ${c_0} = - 2$,
and therefore 
\begin{equation}
\sin \theta {{dB}\over {d\theta}} = E (1 - 2 \cos \theta - 3 {\cos ^2} \theta ) \; \; \; .
\label{nullBsoln}
\end{equation}
Now from eqn. (\ref{axidivm}) we obtain
\begin{equation}
{d \over {d\theta}} \left ( {\sin^2} \theta C\right ) = E \sin \theta (1 - 2 \cos \theta - 3 {\cos ^2} \theta ) \; \; \; ,
\end{equation}
for which the solution is 
\begin{equation}
{\sin^2} \theta C = E ( - \cos \theta - {\sin ^2} \theta + {\cos ^3} \theta + {c_1} ) \; \; \; ,
\end{equation}
for some constant $c_1$.  Since the left hand side vanishes at $\theta = \pi$ it follows that ${c_1}=0$ and thus
\begin{equation}
C = - E (1 + \cos \theta ) \; \; \; .
\label{nullCsoln}
\end{equation}
Then using eqn. (\ref{memoryformula}) it follows that the displacement due to the null part of the memory is  
\begin{equation}
\Delta {D_a} = {\frac E r} (1+\cos \theta) ({\theta _a}{\theta _b} - {\phi _a}{\phi _b}) {D^b} \; \; \; .
\label{nullmemoryresult}
\end{equation}
Comparing to eqn. (\ref{memoryresult}) we find that to first order in $E/M$ the memory is entirely null memory, as 
asserted in \cite{twmemory}.

We now calculate the ordinary memory.  For this we must calculate $\Delta P$.  Note that before the particle decays, the
metric perturbation is just that of a Schwarzschild metric of mass $M$.  Therefore $P(-\infty)$ is just the $P$ of 
Schwarzschild.  After the decay, and after the null particle has hit null infinity, the metric perturbation is again, that of a Schwarzschild metric, but now with mass $M'$ and boosted with velocity $v$ in the $-z$ direction.  We thus need to calculate the $P$ of both boosted and unboosted Schwarzschild.  In Appendix \ref{Pcalculation}, we will derive the following result: for a particle of energy $\cal E$ moving with velocity $V {\hat z}$, where $|V| < 1$, the quantity $P$ is given by
\begin{equation}
P = {\frac {- 2 {\cal E} {{(1 - {V^2})}^2}} {{(1 - V \cos \theta )}^3}} \; \; \; .
\label{Presult}
\end{equation}  
Before the decay, we have a particle of mass $M$ and zero velocity, so it follows that $P(-\infty)= - 2 M$.  After the decay, it follows from eqns. (\ref{conserveE}-\ref{conserveP}) that the recoiling particle has energy $M-E$ 
and velocity $V = -E/(M-E)$.  It therefore follows from eqn. (\ref{Presult}) that 
\begin{equation}
P(\infty) = {\frac {- 2 M {{\left ( 1 - {\frac {2E} M}\right ) }^2}} {{\left ( 1 - {\frac E M}(1-\cos \theta ) \right ) }^3}} \; \; \; ,
\end{equation}  
and therefore that
\begin{equation}
\Delta P = 2 M \left [ 1 - {\frac {{\left ( 1 - {\frac {2E} M}\right ) }^2} {{\left ( 1 - {\frac E M}(1-\cos \theta ) \right ) }^3}} \right ] \; \; \; .
\label{DeltaP}
\end{equation}
Then computing and subtracting the $\ell=0$ and $\ell=1$ part of eqn. (\ref{DeltaP}) we find that
\begin{equation}
\Delta P - {{(\Delta P)}_{[1]}} = 2 M \left [ 1 - {\frac {{\left ( 1 - {\frac {2E} M}\right ) }^2} {{\left ( 1 - {\frac E M}(1-\cos \theta ) \right ) }^3}} - {\frac E M} (1+3\cos \theta )\right ] \; \; \; .
\label{DeltaPc}
\end{equation}
We are now in a position to explain the agreement of the calculation of \cite{twmemory} with the null memory.  Since that calculation is first order in $E/M$ and agrees with the null memory, it follows that to first order in $E/M$ the ordinary memory must vanish.  However, the ordinary memory is due to the recoiling particle, and we would certainly expect that 
$\Delta P$ of the recoiling particle contains terms that are first order in $E/M$.  Indeed, it follows 
from eqn. (\ref{DeltaP}) that to first order in $E/M$ we have $\Delta P = 2E(1+3\cos \theta)$.  Thus, though to first order 
$\Delta P$ does not vanish, it consists purely of $\ell =0$ and $\ell =1$ parts.  Since those parts do not contribute to the memory, it follows that to first order in $E/M$ the ordinary memory for this process vanishes.      

Now to find the ordinary memory, we must solve eqns. (\ref{axilaplace}-\ref{axidivm}) with $A$ given by the right hand side
of eqn. (\ref{DeltaPc}).  Define the quantities $s$ and $X$ by $s=E/M$ and
\begin{equation}
X = 1-s(1-\cos \theta) \; \; \; .
\label{Xdef}
\end{equation}
Then eqn. (\ref{axilaplace}) becomes
\begin{equation}
{\frac d {d\theta}} \left ( \sin \theta {\frac {dB} {d\theta}} \right ) 
= M \sin \theta \left [ 8 (1-s) - 6 X - 2 {{(1-2s)}^2}{X^{-3}} \right ] \; \; \; .
\end{equation}
Integrating this equation we find
\begin{equation}
\sin \theta {\frac {dB} {d\theta}}  
= - {\frac M s} \left [ 8 (1-s) X - 3 {X^2} + {{(1-2s)}^2}{X^{-2}} + {c_0} \right ] \; \; \; ,
\label{firstintegral}
\end{equation}
where $c_0$ is a constant.  This constant must be chosen so that the right hand side of eqn.(\ref{firstintegral})
vanishes at $\theta =0$, which corresponds to $X=1$.  It then follows that
\begin{equation}
{c_0} = -8(1-s) + 3 - {{(1-2s)}^2} \; \; \; .
\label{c0}
\end{equation}
Eqn. (\ref{axidivm}) then becomes
\begin{equation}
{\frac d {d\theta}} ({\sin^2}\theta C) = - {\frac {M \sin \theta} s} \left [ 8 (1-s) X - 3 {X^2} + {{(1-2s)}^2}{X^{-2}} + {c_0} \right ] \; \; \; .
\end{equation}
Integrating this equation we find
\begin{equation}
{\sin^2}\theta C = {\frac M {s^2}} \left [ 4(1-s){X^2} - {X^3} - {{(1-2s)}^2}{X^{-1}} + {c_0}X +{c_1}\right ] \; \; \; ,
\label{secondintegral}
\end{equation}
where $c_1$ is a constant.
The right hand side of eqn. (\ref{secondintegral}) must vanish at $\theta =0$, which yields
\begin{equation}
{c_1} = -4(1-s) + 1 + {{(1-2s)}^2} - {c_0} \; \; \; .
\label{c1}
\end{equation}
Using eqns. (\ref{c0}) and (\ref{c1}) in eqn. (\ref{secondintegral}), some straightforward algebra yields
\begin{equation}
{\sin ^2} \theta C = - {\frac M {s^2}} {X^{-1}} {{(X-1)}^2} {{(X-[1-2s])}^2} \; \; \; .
\label{simpleC}
\end{equation}
Then using eqn. (\ref{Xdef}) in eqn. (\ref{simpleC}) we obtain
\begin{equation}
C = - \left ( {\frac {E^2} M}\right ) {\frac {{\sin^2}\theta} {1 - (E/M)(1-\cos \theta)}} \; \; \; .
\label{finalC}
\end{equation}
Then using eqn. (\ref{memoryformula}) it follows that the displacement due to the ordinary part of the memory is  
\begin{equation}
\Delta {D_a} = \left ( {\frac {E^2} {Mr}}\right ) {\frac {{\sin ^2}\theta} {1 - (E/M)(1-\cos \theta )}} ({\theta _a}{\theta _b} - {\phi _a}{\phi _b}) {D^b} \; \; \; .
\label{ordinarymemoryresult}
\end{equation}
Adding the null memory displacement of eqn. (\ref{nullmemoryresult}) to the ordinary memory displacement of 
eqn. (\ref{ordinarymemoryresult}) yields the total displacement, which agrees with the result of eqn. (\ref{memoryresult}).

\section{memory of a timelike particle}

We now consider the memory due to the decay of a particle of mass $M$ where both particles produced in the decay are timelike.  The particle moving in the $z$ direction will have energy $E$ and velocity $\beta {\hat z}$ where $0 < \beta < 1$.  The recoil particle will have energy ${\tilde E}$ and velocity ${\tilde \beta} {\hat z}$ where $-1 < {\tilde \beta} < 0$.  Note that
$\tilde E$ and $\tilde \beta $ are not independent quantities: the conservation of energy and momentum in the decay requires
\begin{eqnarray}
M = E + {\tilde E} \; \; \; ,
\label{timelikeEnergy}
\\
0 = E \beta + {\tilde E}{\tilde \beta} \; \; \; ,
\label{timelikeMomentum}
\end{eqnarray}
which yields 
\begin{eqnarray}
{\tilde E} = M - E \; \; \; ,
\label{tildeE}
\\
{\tilde \beta} = {\frac {- \beta E} {M-E}} \; \; \; .
\label{tildebeta}
\end{eqnarray}

First we calculate the memory using the method of \cite{twmemory}.  As in section II, the axisymmetry of the problem means that 
the electric part of the Weyl tensor is of the form in eqn. (\ref{electricWeyl}) with $W$ given by eqn. (\ref{Wcalc}) and 
$R_{txty}$ given by eqn. (\ref{WeylExy1}).  Thus, we only need to calculate the perturbed metric component $h_{tt}$.  Note that the situation is very similar to that of \cite{twmemory}, with the same metric before the decay, and after the decay the null particle and recoiling particle replaced by two timelike particles.  It then follows from eqns. (44) and (45) in \cite{twmemory} that in our case to leading order in $1/r$ we have
\begin{equation}
{h_{tt}} = {\frac k r} + {\frac 1 r} \Theta (u) \left [ - 2 M + {\frac {2 E (1 + {\beta^2})} {1 - \beta \cos \theta}}
+ {\frac {2 {\tilde E} (1 + {{\tilde \beta}^2})} {1 - {\tilde \beta} \cos \theta}} \right ] \; \; \; .
\label{timelikehtt1}
\end{equation}
Then applying eqns. (\ref{timelikeEnergy}-\ref{timelikeMomentum}) to eqn. (\ref{timelikehtt1}) we obtain
\begin{equation}
{h_{tt}} = {\frac k r} + {\frac 1 r} \Theta (u) {\frac {2 E \beta (\beta - {\tilde \beta})(1 + {\cos ^2} \theta )}
{(1 - \beta \cos \theta )(1-{\tilde \beta}\cos \theta )}} \; \; \; .
\label{timelikehtt2}
\end{equation}  
Then using eqn. (\ref{timelikehtt2}) in eqn. (\ref{WeylExy1}) we obtain
\begin{equation}
{R_{txty}} = {{-1}\over r} {\Theta ''} (u) {\frac {E \beta (\beta - {\tilde \beta})(1 + {\cos ^2} \theta )}
{(1 - \beta \cos \theta )(1-{\tilde \beta}\cos \theta )}}  {\sin ^2} \theta \cos \phi \sin \phi \; \; \; .
\label{timelikeWeylExy2}
\end{equation}
Now using eqn. (\ref{timelikeWeylExy2}) in eqn. (\ref{Wcalc}) we obtain
\begin{equation}
W = {{- 1}\over r} {\Theta ''} (u)  {\frac {E \beta (\beta - {\tilde \beta}){\sin ^2} \theta }
{(1 - \beta \cos \theta )(1-{\tilde \beta}\cos \theta )}} \; \; \; .
\label{timelikeWresult}
\end{equation}
Finally, using eqn. (\ref{timelikeWresult}) in eqn. (\ref{geodev}), integrating twice with respect to time, and using eqn. (\ref{tildebeta}) to elimnate $\tilde \beta$ we find that the
net change in separation $\Delta {D_a}$ is given by
\begin{equation}
\Delta {D_a} = {\frac {E {\beta ^2} {\sin ^2} \theta} {r (1 - \beta \cos \theta ) (1-(E/M)(1 - \beta \cos \theta))}} ({\theta _a}{\theta _b} - {\phi _a}{\phi _b}) {D^b} \; \; \; .
\label{timelikememory}
\end{equation}

In order to get some insight into the relation between ordinary memory and null memory, we calculate the memory of the timelike decay again, but this time using the method of \cite{lydiaandme}.
It follows from eqn. (\ref{Presult}) that for this decay process we have
\begin{equation} 
\Delta P = 2 M - {\frac {2E {{(1-{\beta ^2})}^2}} {{(1-\beta \cos \theta)}^3}} 
- {\frac {2{\tilde E} {{(1-{{\tilde \beta} ^2})}^2}} {{(1-{\tilde \beta} \cos \theta)}^3}} \; \; \; .
\label{timelikeDeltaP}
\end{equation}
Because in this case the entire memory is ordinary memory, it follows that the $\ell=0$ and $\ell=1$ parts of $\Delta P$ vanish, so there is no need to perform a subtraction of these parts.
To find the memory, we need to solve eqns. (\ref{axilaplace}-\ref{axidivm}) with $A$ given by the right hand side of 
eqn. (\ref{timelikeDeltaP}).  Integrating eqn. (\ref{axilaplace}) we obtain
\begin{equation}
\sin \theta {\frac {dB} {d\theta}} = - 2 M \cos \theta + {\frac {E {{(1-{\beta ^2})}^2}} 
{\beta{{(1-\beta \cos \theta)}^2}}} 
+ {\frac {{\tilde E} {{(1-{{\tilde \beta} ^2})}^2}} {{\tilde \beta}{{(1-{\tilde \beta} \cos \theta)}^2}}}
+ {c_0} \; \; \; .
\end{equation}
The constant of integration $c_0$ is fixed by demanding that the right hand side vanish at $\theta =0$ which, using eqns. 
(\ref{timelikeEnergy}-\ref{timelikeMomentum}) yields
\begin{equation}
{c_0} = -1 \left ( {\frac E \beta} + {\frac {\tilde E} {\tilde \beta}} \right ) \; \; \; .
\label{timelikec0}\end{equation}
Equation (\ref{axidivm}) then becomes
\begin{equation}
{\frac d {d\theta}} ({\sin ^2} \theta C ) = \sin \theta \left [ - 2 M \cos \theta + {\frac {E {{(1-{\beta ^2})}^2}} 
{\beta{{(1-\beta \cos \theta)}^2}}} 
+ {\frac {{\tilde E} {{(1-{{\tilde \beta} ^2})}^2}} {{\tilde \beta}{{(1-{\tilde \beta} \cos \theta)}^2}}}
+ {c_0} \right ] \; \; \; ,
\end{equation}
from which we obtain
\begin{equation}
{\sin ^2} \theta C = M {\cos ^2} \theta - {\frac {E {{(1-{\beta ^2})}^2}} {{\beta ^2}(1-\beta \cos \theta)}}
- {\frac {{\tilde E} {{(1-{{\tilde \beta} ^2})}^2}} {{{\tilde \beta} ^2}(1-{\tilde \beta} \cos \theta)}}
- {c_0} \cos \theta + {c_1} \; \; \; .
\label{timelikeCprelim}
\end{equation}
The constant of integration $c_1$ is fixed by demanding that the right hand side vanish at $\theta =0$ which, using eqns.
(\ref{timelikeEnergy}), (\ref{timelikeMomentum}), and (\ref{timelikec0}) yields
\begin{equation}
{c_1} = {\frac E {\beta ^2}} + {\frac {\tilde E} {{\tilde \beta}^2}} - 2 M \; \; \; .
\label{timelikec1}
\end{equation}
Finally, applying eqns. (\ref{timelikeEnergy}), (\ref{timelikeMomentum}), (\ref{timelikec0}), and (\ref{timelikec1}) 
to eqn. (\ref{timelikeCprelim}) some straightforward algebra yields 
\begin{equation}
C = {\frac {- E {\beta ^2} {\sin ^2} \theta} {(1 - \beta \cos \theta ) (1-(E/M)(1 - \beta \cos \theta))}} \; \; \; .
\label{timelikeCfinal}
\end{equation}
Then using eqn. (\ref{memoryformula}) it follows that the displacement is given by eqn. (\ref{timelikememory})

We now consider the null limit of the timelike decay, that is we consider at fixed $E$ the limit as $\beta \to 1$.  First note that in the limit as $\beta \to 1$ eqn. (\ref{timelikememory}) goes to eqn. (\ref{memoryresult}).  That is, as the timelike particle approaches the speed of light the memory produced by the timelike decay approaches the memory produced by the null decay.  Though this is certainly what we would expect, we now consider how to reconcile this limit with our picture of the two types of gravitational wave memory.  The null decay has both ordinary memory sourced by $\Delta P$ and null memory sourced 
by $F$.  The timelike decay has only ordinary memory.  Thus, since the memory of the timelike decay approaches that of the null decay in the limit as $\beta \to 1$, it follows that some piece of $\Delta P$ must mimic the $-8\pi F$ of the null particle.  In particular, define $\Delta {P_E}$ to be the middle term on the right hand side of eqn. (\ref{timelikeDeltaP}).  That is   
\begin{equation}
\Delta {P_E} = - {\frac {2E {{(1-{\beta ^2})}^2}} {{(1-\beta \cos \theta)}^3}} \; \; \; .
\label{DeltaPE}
\end{equation}
In physical terms, one can think of $\Delta {P_E}$ as the contribution of the particle of energy $E$ to the source of the memory.  It follows from eqn. (\ref{DeltaPE}) that for $\theta \ne 0$ we have ${\lim _{\beta \to 1}} \Delta {P_E} = 0$ and
that for all $\beta < 1$ we have $\int \Delta {P_E} d \Omega = - 8 \pi E $ where the integral is over the unit two-sphere and $d \Omega$ is the usual volume element.  It then follows that in a distributional sense we have 
${\lim _{\beta \to 1}} \Delta {P_E} = - 8 \pi E \delta$.  Thus, in the limit as the timelike particle becomes null the 
$\Delta P$ of the timelike particle becomes the $- 8 \pi F$ of the null particle.

\section{conclusions}

Our calculations of the memory due to particle decay provide a simple and explicit example of two types of gravitational wave memory. The null memory is associated with the angular distribution of energy radiated to null infinity.  The ordinary memory is associated with the change in the quantity $P$ which has to do with the asymptotic state of the matter that does not get to null infinity.  However, as emphasized in \cite{twmemory}, one should not think of the memory as being ``caused'' by the radiation of energy to null infinity (nor by the change in $P$).  Rather, it is the decay process itself which creates gravitational waves that give rise to the memory.  That same decay process also results in energy radiated to null infinity and a change in the quantity $P$.  Nonetheless, the memory is associated with energy radiated to null infinity and change in $P$ in the sense that knowledge of these quantities alone is sufficient to calculate the memory.

Ordinary and null memory are distinct just as timelike particles differ from null particles.  Nonetheless, just as a timelike particle with high velocity mimics a null particle, so the ordinary memory can mimic the null memory.  This comes about because for high velocity the $C_{trtr}$ component of the curvature is strongly peaked in the forward direction and thus mimics the energy flux of a null particle.  

Our use of the point particle idealization limits our results to linearized gravity, since point particles do not make sense in the full nonlinear theory of general relativity.\cite{bobandjennie}  Nonetheless, we expect that our conclusions on the nature of gravitational wave memory continue to hold in the full theory.

\section*{Acknowledgements}

A. T. and R. M. W. were supported by NSF Grant No. PHY-1202718 to The University of Chicago.  L. B. was supported by NSF grant No. DMS-1253149 to The University of Michigan.  D. G. was supported by NSF Grant No. PHY-1205202 to Oakland University.     

\appendix

\section{distributional memory solution}
\label{distribution}

We now verify that the solution found in section II for the memory of a null particle is a distributional solution.  For
$\Phi_2$ to be a distributional solution of eqn. (\ref{laplacenull}) means that 
for any smooth function $g$ on the two-sphere we have
\begin{equation}
\int d \Omega \left [ {\Phi _2} {D_A}{D^A} g + 8 \pi (F - {F_{[1]}}) g \right ] = 0 \; \; \; ,
\label{laplacedistro}
\end{equation}
where the integral is over the two-sphere with $d\Omega$ the usual two-sphere volume element.  Thus, we must evaluate the left hand side of eqn. (\ref{laplacedistro}) with $\Phi _2$ equal to the $B$ specified in eqn. (\ref{nullBsoln}) 
and $F-{F_{[1]}}$ given in eqn. (\ref{nullsource}), and 
$g$ an arbitrary smooth function.  If the result is zero, then the solution is a distributional solution.  We have
\begin{eqnarray}
\int d \Omega \left [ B {D_A}{D^A} g + 2 E (4 \pi \delta - (1 + 3 \cos \theta )) g \right ] 
\nonumber
\\
= 8 \pi E g{|_{\theta =0}} + {\lim _{\epsilon \to 0}} {\int _{\theta > \epsilon}} d \Omega \left [ B {D_A}{D^A} g 
- 2 E (1 + 3 \cos \theta ) g \right ] 
\nonumber
\\
= 8 \pi E g{|_{\theta =0}} + {\lim _{\epsilon \to 0}} {\int _{\theta > \epsilon}} d \Omega {D_A}(B {D^A} g - g {D^A}B)
\nonumber 
\\
+ {\lim _{\epsilon \to 0}} {\int _{\theta > \epsilon}} d \Omega g \left [ {D_A}{D^A} B 
- 2 E (1 + 3 \cos \theta )  \right ]
\nonumber
\\
= 8 \pi E g{|_{\theta =0}} + {\lim _{\epsilon \to 0}}  {{\left [ (- 2 \pi \sin \theta ) \left ( B {\frac {\partial g} {\partial \theta}} - g {\frac {dB} {d\theta}} \right ) \right ]}_{\theta = \epsilon}}
\nonumber 
\\
+ {\lim _{\epsilon \to 0}} {\int _{\theta > \epsilon}} d \Omega g \left [ {\frac 1 {\sin \theta}} {\frac d {d \theta}}
\left ( \sin \theta {\frac {dB} {d\theta}}\right )  
- 2 E (1 + 3 \cos \theta )  \right ]
\nonumber
\\
= ( 2 \pi g{|_{\theta =0}} ) \left [ 4 E + {\lim _{\theta \to 0}} \sin \theta {\frac {dB} {d\theta}}  \right ]
\nonumber 
\\
+ {\lim _{\epsilon \to 0}} {\int _{\theta > \epsilon}} d \Omega g \left [ {\frac 1 {\sin \theta}} {\frac d {d \theta}}
\left ( E ( 1 - 2 \cos \theta - 3 {\cos ^2} \theta ) \right )  
- 2 E (1 + 3 \cos \theta )  \right ]
\nonumber
\\
= ( 2 \pi g{|_{\theta =0}} ) \left [ 4 E + {\lim _{\theta \to 0}}  E ( 1 - 2 \cos \theta - 3 {\cos ^2} \theta ) \right ] = 0 \; \; \; .
\end{eqnarray}
Therefore the $B$ specified by eqn. (\ref{nullBsoln}) is a distributional solution of eqn. (\ref{laplacenull}) 
with the $F-{F_{[1]}}$ given in eqn. (\ref{nullsource}).  

For eqn. (\ref{divmnull}) to be satisfied in a distributional sense means 
that for any smooth vector field $V^A$ on the two-sphere we have
\begin{equation}
\int d \Omega \left [ {m_{2AB}}{D^B}{V^A} - {\Phi _2}{D_A}{V^A} \right ] = 0 \; \; \; .
\label{divmdistro}
\end{equation}
Thus we must evaluate the left hand side of eqn. (\ref{divmdistro}) with 
$m_{2AB}$ given by the expression in eqn. (\ref{aximemory}) with $C$ given in eqn. (\ref{nullCsoln})
and with $\Phi_2$ equal to the $B$ specified in eqn. (\ref{nullBsoln}) and with $V^A$ an arbitrary smooth vector field.  
If the result is zero, then eqn. (\ref{divmnull}) is satisfied in a distributional sense.  We have  
\begin{eqnarray}
\int d \Omega \left [ C ({\theta _A}{\theta _B} - {\phi _A}{\phi _B}) {D^B}{V^A} - B {D_A}{V^A} \right ] 
\nonumber 
\\
= {\lim _{\epsilon \to 0}} {\int _{\theta > \epsilon}} d \Omega \left [ C ({\theta _A}{\theta _B} - {\phi _A}{\phi _B}) {D^B}{V^A} - B {D_A}{V^A} \right ]
\nonumber 
\\
= {\lim _{\epsilon \to 0}} {\int _{\theta > \epsilon}} d \Omega {D^B} \left [ C ({\theta _A}{\theta _B} - {\phi _A}{\phi _B}) {V^A} - B {V_B} \right ]
\nonumber 
\\
+ {\lim _{\epsilon \to 0}} {\int _{\theta > \epsilon}} d \Omega \left [ - {V^A}{D^B} (C ({\theta _A}{\theta _B} - {\phi _A}{\phi _B}))  + {V^B}{D_B} B  \right ]
\nonumber
\\
= {\lim _{\theta \to 0}} 2 \pi \sin \theta {V^\theta} (B-C) 
\nonumber
\\
+ {\lim _{\epsilon \to 0}} {\int _{\theta > \epsilon}} d \Omega \, {V^\theta} \left [ {\frac {dB} {d\theta}} - 
{\frac {dC} {d\theta}} - 2 \cot \theta C \right ]
\nonumber
\\
= {\lim _{\epsilon \to 0}} {\int _{\theta > \epsilon}} d \Omega \, {\frac {V^\theta} {\sin \theta}} \left [ 
\sin \theta {\frac {dB} {d\theta}} - \sin \theta
{\frac {dC} {d\theta}} - 2 \cos \theta C \right ]
\nonumber
\\
= {\lim _{\epsilon \to 0}} {\int _{\theta > \epsilon}} d \Omega \, {\frac {V^\theta} {\sin \theta}} \left [ 
E (1 - 2 \cos \theta - 3 {\cos ^2} \theta) - E {\sin^2}\theta + 2 E \cos \theta (1 + \cos \theta) \right ] = 0 \; \; \; .
\end{eqnarray}
Therefore the $B$ and $C$ given respectively by eqns. (\ref{nullBsoln}) and (\ref{nullCsoln}) provide 
a distributional solution of eqn. (\ref{divmnull}).

\section{Calculation of P for boosted Schwarschild perturbation}
\label{Pcalculation}

Associated with the usual spherical coordinates $(t,r,\theta,\phi)$ there is the usual orthonormal tetrad
$({t^a},{r^a},{\theta^a},{\phi^a})$.  Introduce the null tetrad $({\ell ^a},{n^a},{m^a},{{\bar m}^a})$ given by
\begin{eqnarray}
{\ell ^a} = {\frac 1 {\sqrt 2}} ({t^a}+{r^a}) \; \; \; ,
\\
{n^a} = {\frac 1 {\sqrt 2}} ({t^a}-{r^a}) \; \; \; ,
\\
{m^a} = {\frac 1 {\sqrt 2}} ({\theta ^a} + i {\phi ^a}) \; \; \; ,
\\
{{\bar m}^a} = {\frac 1 {\sqrt 2}} ({\theta ^a} - i {\phi ^a}) \; \; \; .
\end{eqnarray}
The Schwarzschild metric of mass $M$ has 
(to first order in perturbation of the flat metric $\eta _{ab}$) the Weyl tensor\cite{Exact}
\begin{equation}
{C_{abcd}} = - {\frac M {r^3}} \left ( {\eta _{ac}}{\eta _{bd}} - {\eta _{ad}}{\eta _{bc}} + 12 {\ell _{[a}}{n_{b]}}
{\ell _{[c}}{n_{d]}} + 12 {m_{[a}}{{\bar m}_{b]}}{m_{[c}}{{\bar m}_{d]}} \right ) \; \; \; .
\end{equation}
Now consider a mass $M$ moving with velocity $V {\hat z}$.  Then the mass is at rest in the coordinate system 
$({t'},{x'},{y'},{z'})$ where 
\begin{eqnarray}
{t'} = \gamma (t- V z) \; \; \; ,
\label{Lorentzt}
\\
{z'} = \gamma (z - V t) \; \; \; ,
\label{Lorentzz}
\end{eqnarray}
where $\gamma = {{(1-{V^2})}^{-1/2}}$ and the $x$ and $y$ coordinates are unchanged.  
The Weyl tensor then takes the form
\begin{equation}
{C_{abcd}} = - {\frac M {{r'}^3}} \left ( {\eta _{ac}}{\eta _{bd}} - {\eta _{ad}}{\eta _{bc}} + 12 
{{\ell'} _{[a}}{{n'}_{b]}}{{\ell'} _{[c}}{{n'}_{d]}} + 12 {{m'}_{[a}}{{{\bar m}'}_{b]}}{{m'}_{[c}}{{{\bar m}'}_{d]}} \right ) \; \; \; .
\end{equation}
We would like to express the Weyl tensor of the moving mass in terms of the coordinates and null tetrad of the stationary observer.  Since we are interested in quantities at null infinity, we will work only to leading order in $1/r$. 
From eqns. (\ref{Lorentzt}-\ref{Lorentzz}) it follows that 
\begin{eqnarray}
{r'} = r \gamma (1 - V \cos \theta) \; \; \; ,
\label{rprime}
\\
{u'} = {\frac u {\gamma (1 - V \cos \theta )}} \; \; \; .
\label{uprime}
\end{eqnarray}
From eqn. (\ref{uprime}) we obtain
\begin{equation}
{{\ell '}_a} = {\frac 1 {\gamma (1 - V \cos \theta)}} {\ell _a} \; \; \; .
\label{lprime}
\end{equation}
Then from eqns. (\ref{Lorentzt}) and (\ref{lprime}) we find
\begin{equation}
{{n'}_a} = \gamma \left [ {\frac {{V^2} {\sin ^2} \theta} {1 - V \cos \theta}} {\ell _a} + (1 - V \cos \theta ){n_a}
- V \sin \theta ({m_a}+{{\bar m}_a} ) \right ] \; \; \; .
\end{equation}
Finally using the fact that $\phi$ and $r \sin \theta$ are unchanged by the Lorentz transformation, we obtain
\begin{equation}
{{m'}_a} = {m_a} - {\frac {V \sin \theta} {1 - V \cos \theta}} {\ell _a} \; \; \; .
\end{equation}
(The complex conjugate of this equation gives the transformation for ${\bar m}_a$).  
We then find that the quantity $P$ is given by
\begin{eqnarray}
P = {r^3}{C_{trtr}}
\nonumber
\\
= {r^3} {\ell ^a}{n^b}{\ell ^c}{n^d}{C_{abcd}}
\nonumber
\\
= - M {{\left ( {\frac r {r'}} \right ) }^3} \left ( - 1 + 3 {{({\ell ^a}{{n'}_a})}^2} {{({n ^b}{{\ell '}_b})}^2} \right )
\\
= {\frac {- 2 M} {{\gamma ^3} {{(1- V \cos \theta )}^3}}} \; \; \; .
\end{eqnarray}
However, $M \gamma$ is the energy $\cal E$ of the particle, and ${\gamma ^{-4}} = {{(1-{V^2})}^2}$.  We therefore obtain
\begin{equation}
P = {\frac {- 2 {\cal E}{{(1-{V^2})}^2}} {{(1- V \cos \theta )}^3} } \; \; \; .
\end{equation}


\begin{thebibliography}{}

\bibitem{zeldovich}
Ya.B. Zeldovich and A.G. Polnarev, Sov. Astron. {\bf 18}, 17 (1974)

\bibitem{christodoulou}
D. Christodoulou, {\it Phys. Rev. Lett.}  {\bf 67}, 1486 (1991)

\bibitem{globalCK}
D. Christodoulou and S. Klainerman, 
{\it The global nonlinear stability of the Minkowski space.}
Princeton Math.Series {\bf 41}. 
Princeton University Press. Princeton. NJ. (1993). 


\bibitem{zipser}
N. Zipser.  
        {\it Extensions of the Stability Theorem of the Minkowski Space
in General Relativity. - Solutions of the Einstein-Maxwell Equations.}
        AMS-IP. Studies in Advanced Mathematics. Cambridge. MA. (2009).

\bibitem{lydia1}
 L. Bieri, P. Chen, S.-T. Yau, 
Advances in Theoretical and Mathematical Physics, {\bf 15}, 4, (2011). 

\bibitem{lydia2}
L. Bieri, P. Chen, S.-T. Yau. 
Class.Quantum Grav. {\bf 29}, 21, (2012). 

\bibitem{epstein}
R. Epstein, Astrophys. J. {\bf 223}, 1037 (1978)

\bibitem{turner}
M. Turner, Astrophys. J. {\bf 216}, 610 (1977)

\bibitem{nullfluid}
L. Bieri and D. Garfinkle, Annales Henri Poincar\'e, 2014 DOI 10.1007/s00023-014-0329-1  

\bibitem{lydiaandme}
L. Bieri and D. Garfinkle, Phys. Rev. D {\bf 89}, 084039 (2014)

\bibitem{twmemory}
A. Tolish and R.M. Wald, Phys. Rev. D {\bf 89}, 064008 (2014)

\bibitem{Exact}
H. Stephani, D. Kramer, M. MacCallum, C. Hoenselaers, and E. Herlt, (2003) {\it Exact solutions of Einstein's Field Equations} second edition p 52 (Cambridge University Press, Cambridge)

\bibitem{bobandjennie}
R. Geroch and J. Traschen, Phys. Rev. D {\bf 36}, 1017 (1987)

\end{thebibliography}
\end{document}